\PassOptionsToPackage{dvipsnames}{xcolor}
\documentclass[5p,authoryear]{elsarticle}
\journal{Eur.~J.~Mech.~A/Solids}
\usepackage{amsmath,amssymb,mathtools}
\usepackage[T1]{fontenc} % To get underscores in doi's properly
\usepackage{lmodern}     % To get the usual LaTeX font, but with T1 font encoding.
\normalfont\DeclareFontShape{T1}{lmr}{bx}{sc} { <-> ssub * cmr/bx/sc }{} % Needed to get bold small caps.
\usepackage{hyperref}
\usepackage[dvipsnames]{xcolor}
\usepackage{graphicx}

\hypersetup{colorlinks,%
    citecolor=blue,%
    filecolor=blue,%
    linkcolor=blue,%
    urlcolor=blue}

\usepackage{tikz}
\usetikzlibrary{calc, patterns, angles, quotes, arrows, arrows.meta, decorations.pathmorphing, math, shapes, bending, positioning, decorations.text, decorations.pathreplacing, fadings}

\newcommand{\pdr}[2]{\frac{\partial {#1}}{\partial {#2}}} % \ptl: partial derivative
\newcommand{\intd}{\mathrm{d}} % d used in derivatives
\newcommand{\UA}{\hat U_{\!\mathrm{A}}}
\newcommand{\hA}{\hat h_{\!\:\!\mathrm{A}}}
\newcommand{\bA}{\hat b_{\!\:\!\mathrm{A}}}
\newcommand{\nhA}{h_{\!\:\!\mathrm{A}}}
\newcommand{\nbA}{b_{\!\:\!\mathrm{A}}}
\newcommand{\vect}[1]{\boldsymbol{#1}}
\newcommand{\bnabla}{\vect{\nabla}}
\newcommand{\Abaqus}{\textsc{Abaqus}} % Typesetting for ABAQUS
\newcommand{\Matlab}{\textsc{Matlab}} % Typesetting for MATLAB

\begin{document}
\begin{frontmatter}

\title{\textbf{Modelling Lateral Spread in Wire Flat Rolling}}

\author[wmi,wmg]{Mozhdeh~Erfanian}\ead{Mozhdeh.Erfanian.1@warwick.ac.uk}
\author[wmg]{Carl~D.~Slater}\ead{C.D.Slater@warwick.ac.uk}
\author[wmi,wmg]{Edward~J.~Brambley\corref{cor1}}\ead{E.J.Brambley@warwick.ac.uk}\cortext[cor1]{Corresponding Author}

 \affiliation[wmi]{organization={Mathematics Institute},
             addressline={University of Warwick},
             city={Coventry},
             postcode={CV4 7AL},
             country={UK}}

  \affiliation[wmg]{organization={WMG},
             addressline={University of Warwick},
             city={Coventry},
             postcode={CV4 7AL},
             country={UK}}

%\date{}
\def\elspublication{Published in European Journal of Mechanics --- A/Solids, volume~117, 106027 (2026), \href{https://doi.org/10.1016/j.euromechsol.2026.106027}{doi:10.1016/j.euromechsol.2026.106027}}

\begin{abstract}%
A mathematical model for wire rolling is developed, focusing on predicting the lateral spread.  This provides, for the first time, an analytic model of lateral spread without any fitting parameters. The model is derived directly from the governing equations, assuming a rigid, perfectly plastic material and exploiting the thinness of the wire (in thickness and width) relative to the roller size.  Results are compared against experiments performed on stainless steel wire using $100\,\mathrm{mm}$ diameter rolls, demonstrating accurate predictions of lateral spread across a wide range of wire diameters ($2.96\,\mathrm{mm}$--$7.96\,\mathrm{mm}$) and reduction ratios (20\%--60\%), all without the need for fitting parameters.  Since the model requires only seconds to compute, the model's valid range is explored for varying roll diameter, wire diameter, and reduction ratio, and their effects on the resulting lateral spread characterized. The model can serve as a robust tool for validating FE results, guiding process design, and laying the foundation for future improved models.
\Matlab\ code to evaluate the model is provided in the supplementary material.
\end{abstract}

\end{frontmatter}

%%%%%%%%%%%%%%%%%%%%%%%%%%%%%%%%%%%%%%%%%%%%%%%%
\section{Introduction}
\begin{figure}
    \centering
    \includegraphics[width=1\linewidth,height=0.3\textheight,keepaspectratio]{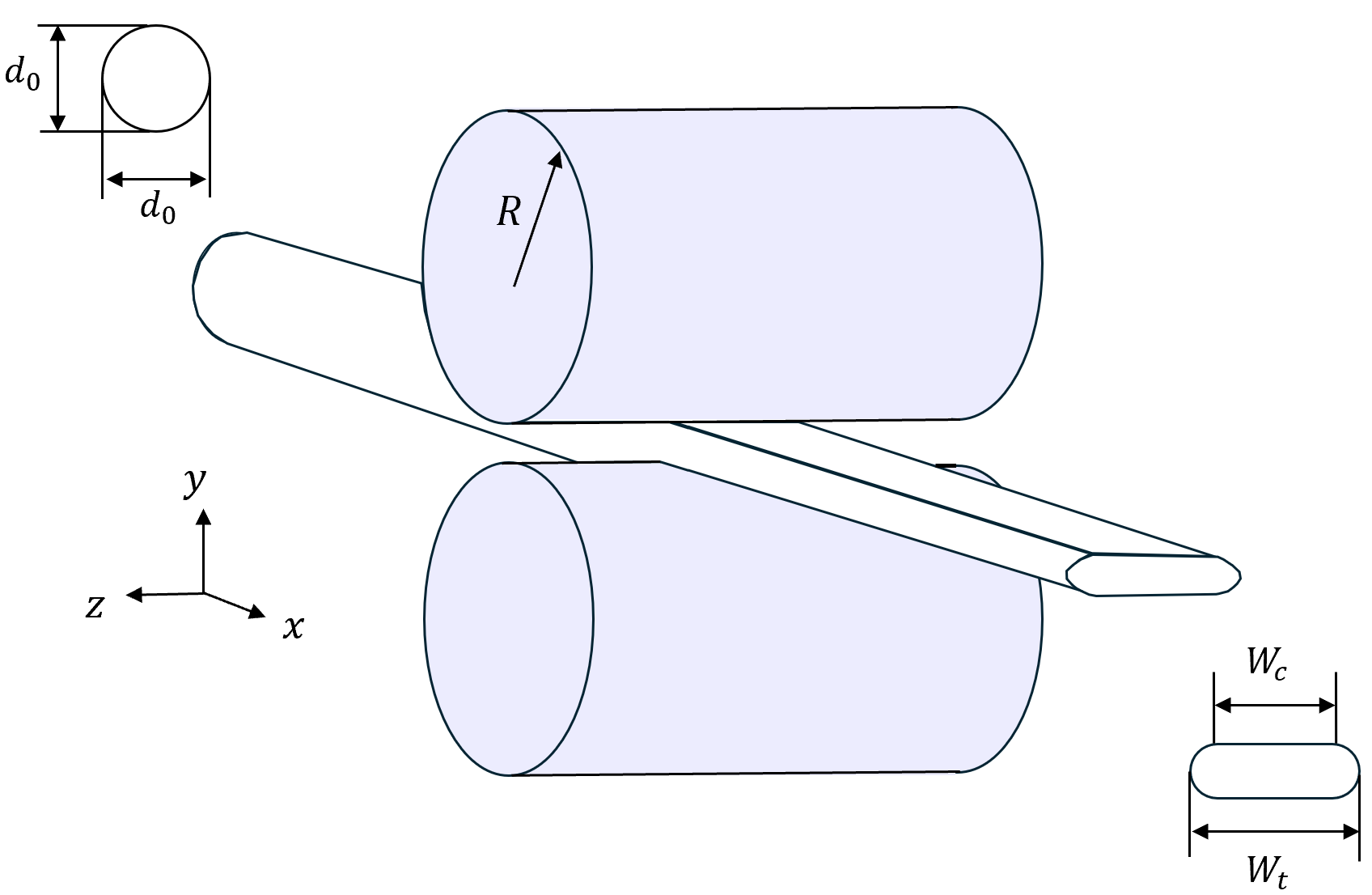}
    \caption{A diagram of wire flat rolling process; the wire initially has a circular cross-section with diameter $d_0$ and is flattened to a barrel shape cross section with the lateral spread $W_t$ and the contact width $W_c$.  Adapted from Figure~1 of~\citet{carlsson1998contact}.}
    \label{fig:wirerolling}
\end{figure}

Applications including sawblades, springs, piston rings, and transformers depend critically on flattened wires~\citep{utsunomiya2001three}. These wires are usually made by a flat rolling technique, where a wire with a circular cross-section is cold rolled between cylindrical rolls --- sometimes through several passes --- to achieve a particular width and thickness (as illustrated in figure~\ref{fig:wirerolling}).
Since the wire can both elongate and widen, achieving a final product that closely matches the desired specifications requires an understanding of how the wire deforms, making it essential to predict the lateral spread accurately.

In cold rolling of a sheet, deformation in the lateral direction often remains within the elastic range due to the product’s geometry. However, this is not the case when an initial width-to-thickness ratio is less than 6, such as in wide strip or plate, or as low as 1 in round wire~\citep{chitkara1966some}. In such cases, the plastic flow in the roll gap at the start of rolling schedule is inherently three-dimensional, complicating the analysis. For round wire, this complexity is further increased by the transformation of a round cross-section into a rectangular shape with bulged edges during the first pass.

Although research exists on the modelling of lateral spread in thick plate rolling~\citep[e.g.][]{chitkara1966some,lahoti1974hill,oh1975approximate,Nikhaily,Kennedy}, wire flat rolling has received less attention. \Citeauthor{kazeminezhad2008error} (\citeyear{kazeminezhad2005width,kazeminezhad2006pressure,kazeminezhad2008error}) published a series of papers studying different parameters in wire rolling. \Citet{kazeminezhad2005width} developed a relationship for the width of the contact area between the rolls and the wire. Their equation was based on experimental and FE observations of the formation of X-shaped shear bands in the wire cross-section during the rolling process \citep{Semiatin1984FormabilityAW,pesin2002mathematical}, along with the assumption that as the height reduction increases, the shear bands rotate while maintaining a constant length.
\Citet{kazeminezhad2006pressure} then applied this equation alongside the slab method, leading to a pressure hill distribution. The study employs the Tresca yield criterion, assuming the longitudinal stress $\sigma_{xx}$ to be the minimum stress, the vertical stress $\sigma_{yy}$ to be the maximum stress, and the compressive stress in the lateral direction, $\sigma_{zz}$, to lie between them. Although $\sigma_{zz}$ is implicitly included in the yield function, and therefore in calculating the roll pressure, the value of $\sigma_{zz}$ is unknown and must be approximated to find the lateral spread. \Citet{kazeminezhad2005width} therefore assumed that $\sigma_{zz}$ takes a value between plane strain and plane stress, written as $\alpha\sigma_{yy}$, where $\alpha$ is a fitting parameter taking values between 0 and 0.5. Using flow rule equations and setting $\sigma_{xx}$ to be zero at the end of the roll gap, they then wrote 
\begin{equation}\label{eq:kaztotalwidth1}
     \frac{d \varepsilon_{zz}}{- d \varepsilon_{yy}} = \frac{\ln \frac{W_t}{d_0}}{\ln \frac{d_0}{2h_1}}= \frac{1-2\alpha}{2-\alpha},
\end{equation}
where $W_t$ is the lateral spread,  $d_0$ is the initial diameter, and $2h_1$ is the final height of the wire. 
The first use of this formulation has been attributed to \citet{hill1955} for plate rolling and has also been derived by several other authors~\citep[e.g.][]{chitkara1966some,applied1961formula}, who experimentally observed a linear relationship between $\ln(W_t/W_0)$ and $\ln(h_0/h_1)$, where $W_0$ and $h_0$ are initial half width and half height of the slab, respectively. 
The factor $\alpha$ and another factor of linearity must then be found empirically from experiments, with a proviso that each resulting equation only gives reasonable lateral spread prediction within the ranges of conditions for which they were empirically determined~\citep{chitkara1966some}. 
For example, for low and high carbon steel~\citet{kazeminezhad2005width} wrote equation~\eqref{eq:kaztotalwidth1} as 
\begin{equation}\label{eq:kaztotalwidth2}
      \frac{W_t}{d_0} = 1.02 \left(\frac{d_0}{2h_1}\right)^{\!0.45}.
\end{equation}

Among a wide range of empirical formulations suggested for predicting lateral spread~\citep{chitkara1966some}, the one proposed by~\citet{kobayashi1978influence} is often referred to as being reliable for round wire~\citep{utsunomiya2001three,kazeminezhad2008error,vallellano2008analysis}.   \Citet{utsunomiya2001three} and \citet{vallellano2008analysis} attribute the following equation to \citet{kobayashi1978influence}, and we will refer to it here by this name: 
\begin{multline}\label{eq:kobayashi}
    \frac{W_t}{d_0}= 0.7854 \frac{d_0}{2h_1} \!\left(\!1 - 15.8 \!\left(\!1 - \frac{2h_1}{d_0} \right)^{\!2.25} \!\left(\frac{2R}{d_0} \right)^{\!-0.82}  \right)
    \\
    + 0.1426 \left(\frac{2h_1}{d_0} \right),
\end{multline}
where $R$ is the roll radius. However, similar to other empirical equations, this equation will be shown to lose accuracy away form the parameters it was initially developed to model.

Finite element (FE) analysis has provided valuable insights into wire rolling, similar to other metal-forming processes. \citet{vallellano2008analysis} performed a 3D numerical analysis using the \Abaqus\ finite element software to study contact stress distributions, residual stresses, and lateral spread. While 3D FE analysis offers highly detailed information, it is computationally expensive, especially when it is used in design iterations where several simulations are required. Although some studies have investigated FE simulations of wire rolling~\citep{carlsson1998contact,vallellano2008analysis,hwang2021influence}, computational cost data are generally not reported. 
However, to give an indication, a 2D FE simulation of half-of-the-thickness sheet rolling required 15.9 CPU hours for a reliable case (30 elements through the half-thickness) and 0.25 CPU hours for a coarse case (5 elements)~\citep{flanagan2025};
since 3D wire-rolling simulations are considerably more complex, they are expected to be significantly more expensive yet.
In addition to high computational cost, FE results require validation against a reliable reference.
Whether the objective is process design or FE validation, a review of the literature further reveals a notable gap in the availability of models with well-defined, traceable assumptions for the rolling of round wire.

Using similar techniques to the asymptotic mathematical modelling of sheet rolling~\citep{erfanian2025}, here an asymptotic mathematical model is developed for the regime of a small friction coefficient and large length-to-thickness ratio, consistent with the slab method of~\citet{kazeminezhad2006pressure}. 
The further assumption that the width is comparably smaller than the length of wire in the roll gap simplifies the problem to plane-stress, where the stress in $z$ direction is negligible. This allows for finding the lateral spread without the need for fitting parameters or solving a problem in a complicated 3D stress state. 
Even so, the derivation of equations in all directions and the use of asymptotic analysis provide a foundation for further improvements to the model in the future, unlike the slab method, which remains inherently limited.

An outline of the paper is as follows. The mathematical model is explained in Section~\ref{sec:model}, beginning with section~\ref{sec:assumptions} introducing the simplifying assumptions.
The scaling parameters and resulting non-dimensionalised governing equations are given in Section~\ref{sec:governingequs} for a rigid perfectly plastic material, and their solution is derived in Section~\ref{sec:solution}.
Section~\ref{sec:numerics} summarises the solutions and outlines the computational methods used to evaluate them. 
The experimental procedure for obtaining data is detailed in Section~\ref{sec:exp}, which is used to validate the lateral spread predictions of the model in Section~\ref{sec:results_width}, alongside FE data from~\citet{vallellano2008analysis}. 
The Longitudinal spread is compared against experimental data in Section~\ref{sec:length} and
the roll pressure predictions are validated against FE data from~\citet{vallellano2008analysis} and presented in Section~\ref{sec:results_press}.
Having validated the model, a parametric study is presented in Section~\ref{sec:parametric_study}, exploring both the range of validity of the model in Section~\ref{sec:range_validity} and the influence of parameters including friction, reduction, and roll diameter on the lateral spread in Section~\ref{sec:parameters_effect}.
Finally, a discussion and potential directions for future research are provided in Section~\ref{sec:conclusion}.
%%%%%%%%%%%%%%%%%%%%%%%%%%%%%%%%%%%%%%%%%%%%%%%%

\section{Mathematical model}\label{sec:model}

\subsection{Assumptions}\label{sec:assumptions}

We first make two assumptions to help in modelling practical wire rolling: section~\ref{sec:cylindrical} concerns the initial transformation of a cylindrical wire in the first roll-stand, and section~\ref{sec:bulge} concerns the bulged edges of the subsequently rolled wire.  These assumptions allow for a simpler model based on rectangular cross-sections to be developed in section~\ref{sec:governingequs}.

\subsubsection{Circular to rectangular deformation}\label{sec:cylindrical}
\begin{figure*}
    \centering
      \includegraphics[width=1\linewidth,height=0.25\textheight,keepaspectratio]{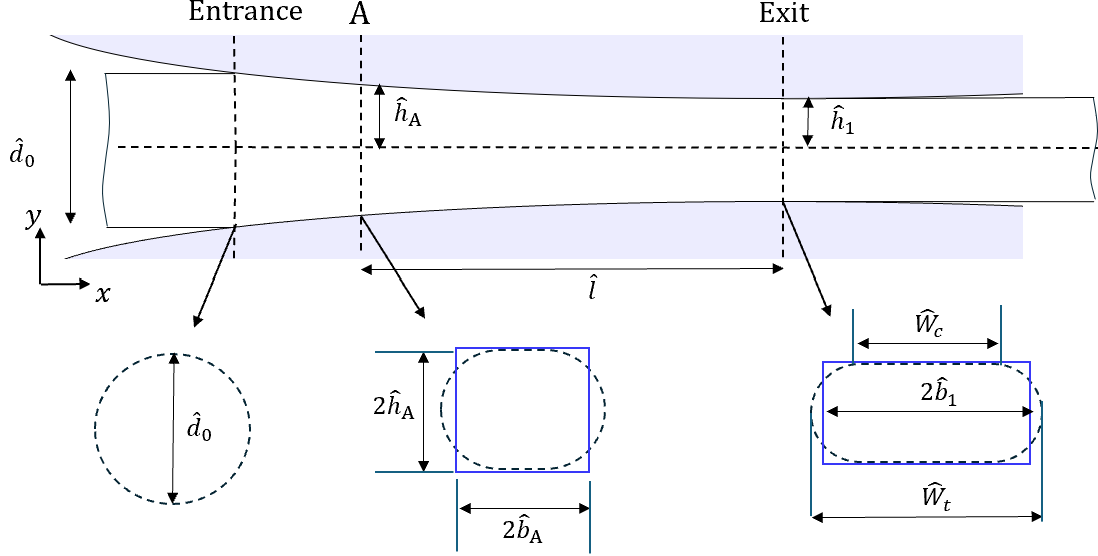}
    \caption{A diagram of the model; the region of interest extends from point A to the roll gap exit. At A, the cross-section is approximated as a square with the same area as the initial round wire, transitioning into a rectangular approximation during rolling with the same area as the real bulged cross section. }
    \label{fig:widthschematics}
\end{figure*}

A schematic diagram of the model is shown in figure~\ref{fig:widthschematics}.
In the first pass of rolling, the round wire undergoes deformation.  Here, we assume that at some point after entry, the initially circular cross-section has transformed into a rectangular shape without changing the cross-sectional area.
This stage is chosen as the beginning of the roll gap in the asymptotic model, and is marked A in figure~\ref{fig:widthschematics}. 
Therefore, instead of solving the equation from the actual roll-gap entrance, where the cross-section is circular, the information is translated to point A in the roll gap, and the equations are solved from this point onwards.
The intuition behind this assumption is that we hypothesise it is much easier for a circular cross-section wire at the beginning of the process --- where only a small region is in contact with the rolls --- to deform into a rectangular shape with a flat top than to elongate. 
It should also be noted that the assumption of an unchanged cross-sectional area between the entrance and point A does not imply that the wire length remains constant, as no assumption about the cross-sectional area is made beyond point~A. All of these assumptions, although physically justifiable, remain approximations that are validated in section~\ref{sec:results} through comparison with experimental results.
The location A is determined by equating the cross-sectional area at this location to that of the initial circular wire.
Thus, if $\hA$ and $\bA$ are the half-thickness and half-width of the wire at location A, and we assume that the initial rectangular cross-section has aspect ratio $\bA/\hA = a$, we may write
\begin{align} \label{eq:h0}
4\hA\bA &= 4a\hA^2 = \frac{\pi}{4} {\hat d_0}\strut^2 &
&\Rightarrow&
\hA &= \sqrt{\frac{\pi}{16a}}\hat d_0
 = \frac{\bA}{a}.
\end{align}
In order to preserve the symmery of the initially cylindrical wire, we will assume that $a=1$ throughout this work unless explicitly stated; this assumption is further justified in \ref{app:rectangular} by comparison with experimental results.  As the quantities will later be non-dimensionalised, and to clearly distinguish between dimensional and non-dimensional parameters, we adopt the convention that, for the remainder of the paper, variables with hats denote dimensional quantities, while unhatted variables represent their dimensionless counterparts. 

Note that this circular-to-square transition is only relevant to the first rolling pass, and for subsequent passes point A is at the roll-gap entrance, $\hA$ and $\bA$ are the half-thickness and half-width of the wire at the entrance, and equation~\eqref{eq:h0} is not needed.

\subsubsection{Bulged edges of rectangular wire}\label{sec:bulge}

In practice, the wire never has a perfectly rectangular cross-section but has bulged edges (see figure~\ref{fig:widthschematics}).  A second assumption is therefore that, as the rolling progresses, a square cross-section is progressively flattened into a rectangular shape, and the bulged edges are approximated as half-circles whilst maintaining the same area as the perfect rectangle (also shown in figure~\ref{fig:widthschematics}).  Consequently,
\begin{align} \label{eq:w_contact}
2 \hat W_c \hat h+\pi \hat h^2 &= 4\hat b \hat h  &
&\Rightarrow&
\hat W_c&=2\hat b - \pi \hat h/2,
\end{align}
where $\hat W_c$ is the width in contact with the roll, and the total lateral spread $\hat W_t$ is given by
\begin{equation} \label{eq:w}
    \hat W_t = \hat W_c + 2 \hat h.
\end{equation}%
This allows a model of the deformation of a rectangular cross-section to be correlated with the real process with a bulged cross-section.
As will be shown later, this set of assumptions allows for the simplest possible model, while having minimal effect on lateral spread prediction.

In the following, we rigorously derive a leading-order asymptotic model for the rolling of a rectangular wire. The ad hoc model previously introduced for the initial transformation of a circular wire into a square cross-section enables this asymptotic formulation to be applied to circular wires as well, which will be validated a posteriori through comparison with FE results.
The governing equations are written for rigid perfectly plastic material rolled with rigid rolls while Coulomb friction is imposed between the rolls and material with a constant friction coefficient. 
The equations are formulated for a regime characterised by a small friction coefficient between the roll and the wire, as well as a wire width and thickness much smaller than its length within the roll gap. The latter condition effectively corresponds to a small wire diameter compared to the roll diameter.

\subsection{Scalings and the non-dimensionalised governing equations} \label{sec:governingequs}

Asymptotic (or perturbation) methods~\citep{hinch} are widely used in applied mechanics, exploiting the smallness of an inherent parameter to obtain approximate analytical/semi-analytical solutions with controllable errors. The first step in applying this technique is to rescale the governing equations and parameters based on the inherent scales of the problem. This process removes physical units and expresses all quantities in a comparable dimensionless form, allowing their relative magnitudes to be evaluated.
The resulting solution can then be redimensionalized using the same scales to provide an engineering solution in physical units.
Here, the horizontal distance is scaled with the characteristic roll-gap length, $\hat \ell$, so that $x = \hat{x}/\hat{\ell}$ gives a dimensionless horizontal coordinate $x$, and the dimensional horizontal position $\hat{x}$ can be recovered from $x$ using $\hat x = \hat\ell x$.  Similarly, the vertical distance is scaled with the initial sheet half-thickness, $\hA$, so that $y = \hat y/\hA$. For circular rolls of radius $\hat R$, the roll-gap length $\hat \ell$ is equal to 
\begin{equation}\label{equ:ell-dimensional}
    \hat \ell = \sqrt{2\hat R(\hA- \hat h_1) - (\hA- \hat h_1)^2},
\end{equation}
where $\hat \ell$ is measured from the point A, as shown in figure~\ref{fig:widthschematics}. 

Similar to the slab method, small values of the friction coefficient and the aspect ratio $\delta = \hA/\hat \ell$ are of interest. The latter can be justified by the practical application of small wire diameters (less than a centimetre) and large roll diameter. 
The roll diameter, wire diameter, and reduction ratio vary significantly depending on the desired reduction, material, and application. Reported values in the literature span a wide range, with roll diameters between 75–1600~$\mathrm{mm}$, wire diameters between 0.5–13~$\mathrm{mm}$, and single-pass reduction ratios between 10$\%$-70$\%$, obtained either experimentally or through FE simulations~\citep{kazeminezhad2005width,vallellano2008analysis,hwang2021influence}. Consequently, the corresponding aspect ratio, $\delta$, also covers a broad range.
For instance, applying a 100~$\mathrm{mm}$ roll diameter to a 7.96~$\mathrm{mm}$ wire diameter with a reduction of 33$\%$ results in an aspect ratio of $\delta = 0.38$, and a larger roll diameter or a smaller wire diameter would further decrease this value.  
The friction coefficient, $\mu$, is typically reported between 0.09–0.3 in wire-rolling studies~\citep{lin2008finite,lee2018development}, suggesting that $\mu$ is of the same order of magnitude as $\delta$.
We also assume that the width, $\hat b(\hat x)$, is of the same order of magnitude as the thickness.  
This is clearly true for the initially cylindrical wire, and becomes less true the wider the wire becomes during rolling.
 We therefore scale the lateral distance $\hat z$ with the initial sheet half-thickness, $\hA$, so that $z = \hat z/\hA$. 
 In this way, with the assumption of the rectangular cross-section without considering barreling, $z$ varies between $-b(x)$ and $b(x)$ where $b(x) = \hat b(\hat x)/ \hA $. 

 The velocity $(\hat u, \hat v, \hat w)$, is scaled with the a priori unknown wire velocity at point A, denoted $\UA$, and with factors of $\delta$ that we will see later are needed to balance incompressibility, giving $(u,\,v,\,w) = (\hat u/\UA, \hat v/\delta\UA, \hat w/\delta\UA)$.
 
 Hydrostatic pressure, $\hat p$, and the normal Cauchy stresses, $\hat \sigma_{xx}$ and $\hat \sigma_{yy}$, and their deviatoric components, $\hat s_{xx}$ and $\hat s_{yy}$, are all scaled with the shear yield stress, $\hat \kappa$. 
The assumption of a small friction coefficient is encoded by setting $\beta = \mu/\delta$, where $\beta $ is a quantity of approximate unit magnitude. Similarly, the shear stress, $\hat \sigma_{xy}$ is scaled with $\delta \hat \kappa$ resulting from small friction. 
To determine the appropriate order of magnitude of stress components in the $z$-direction, we analysed the boundary condition at the lateral edges of the wire. Since these edges are free surfaces, the material is unconstrained in the $z$-direction and must satisfy the traction-free boundary condition
\begin{equation}\label{eq:traction-free}
\vect{\hat\sigma \cdot m} = 0, 
\end{equation}
where $\vect m$ is the unit vector normal to the wire lateral edges. 
By enforcing this physical constraint, we can infer the scaling of $z$-direction stress components relative to $\hat{\kappa}$, and thus determine their order of magnitude in the model.
For a rectangular cross section, the wire lateral edges are given by $\hat{z} - \hat{b}(\hat{x})=0$, so the unit vector normal $\vect{m}$ is given by 
 \begin{equation} \vect{m} = \frac{\hat\bnabla \big(\hat z - \hat b(\hat x)\big)}{\big\|\hat\bnabla \big(\hat z - \hat b(\hat x)\big)\big\|} = \frac{\big(-\delta\intd b/\intd x, 0, 1\big)}{\sqrt{1+\delta^2\big(\intd b/\intd x\big)^2}}\end{equation}%
where $\hat\bnabla = (\partial/\partial\hat x, \partial/\partial\hat y, \partial/\partial\hat z)$, and the factor $\delta$ is due to $\hat b$ being scaled by $\hA$ while $\hat x$ is scaled by $\hat\ell$.
Therefore applying the traction-free boundary condition~\eqref{eq:traction-free} we find
\begin{subequations} \begin{align}
    \delta \frac{\intd b}{\intd x} \sigma_{xx} &= \hat\sigma_{xz}/\hat\kappa &&\text{at } z=b(x), \label{eq:edge_xz}\\
    \delta^2 \frac{\intd b}{\intd x} \sigma_{xy} &= \hat\sigma_{yz}/\hat\kappa &&\text{at } z=b(x), \\
    \delta \frac{\intd b}{\intd x}\hat\sigma_{xz}/\hat\kappa &= \hat\sigma_{zz}/\hat\kappa &&\text{at } z=b(x),
\end{align}\label{eq:edge_noten}\end{subequations}%
where $\intd b / \intd x$, $\sigma_{xx}$, and $\sigma_{xy}$ are written in their known dimensionless forms, and $\hat \sigma_{zz}$, $\hat \sigma_{xz}$, and $\hat \sigma_{yz}$, whose orders of magnitude are thus far unknown, are written in dimensional form.
The set of equations~\eqref{eq:edge_noten} implies that $\hat \sigma_{xz}$ is $O(\delta\hat\kappa)$ and the normal stress in the lateral direction, $\hat \sigma_{zz}$ and the shear stress component $\hat \sigma_{yz}$ are both $O(\delta^2\hat\kappa)$. 
Although this holds at the lateral edges, we assume these orders of magnitude apply throughout the modelling region. 
Consequently, $\hat \sigma_{zz}$ and $\hat \sigma_{yz}$ are scaled with $\delta^2 \hat \kappa$, while $\hat \sigma_{xz}$ is scaled with $\delta \hat \kappa$. 
As will be shown later, this approach leads to the simplest yet consistent solution, where the horizontal velocity and normal stresses depend only on $x$. 
To complete the non-dimensionalisation, the scaling for the plastic multiplier is chosen to balance the horizontal flow equation, $\partial\hat{u}/\partial\hat{x} = \hat\lambda\hat s_{xx}$, giving $\lambda = \hat\kappa\hat\ell\hat \lambda/\UA$. In summary, the original dimensional variables can be obtained from the dimensionless variables using the following rescalings:
\begin{align} 
 \hat{x}&=\hat{\ell} x,&
\hat y&= \hA y,&
\hat z&= \hA z, & \nonumber 
\\
\hat h&= \hA h,&
\hat{b}&=\hA b,&
\hA &= \delta\hat\ell \nonumber 
\\
\hat{\sigma}_{xx}&=\hat{\kappa}\sigma_{xx},&
\hat{\sigma}_{yy}&=\hat{\kappa}\sigma_{yy},&
\hat{\sigma}_{zz}&=\delta^2\hat{\kappa}\sigma_{zz},\nonumber 
\\
\hat{\sigma}_{xy}&=\delta\hat{\kappa}\sigma_{xy},&
\hat{\sigma}_{yz}&=\delta^2\hat{\kappa}\sigma_{yz},&
\hat{\sigma}_{xz}&=\delta\hat{\kappa}\sigma_{xz},\label{eq:scales} \\
\hat{u}&=\UA u, &
\hat{v}&=\delta \UA v,&
\hat{w}&=\delta \UA w,\nonumber 
\\
\hat{\lambda}&=\frac{\UA}{\hat{\ell}\hat{\kappa}} \lambda,&
\hat{p}&=\hat{\kappa}p,&
\mu&= \delta \beta.
\nonumber \end{align} 

We may now use the scalings introduced in equation~\eqref{eq:scales} to write the governing equations in a dimensionless rescaled form.  For example, the vertical force balance is written as
\begin{align}
0 &= \frac{\partial\hat\sigma_{xy}}{\partial\hat x}
+ \frac{\partial\hat\sigma_{yy}}{\partial\hat y}
+ \frac{\partial\hat\sigma_{yz}}{\partial\hat z}
\notag\\&=
\frac{\partial(\delta\hat{\kappa}\sigma_{xy})}{\partial(\hat\ell x)}
+ \frac{\partial(\hat{\kappa}\sigma_{yy})}{\partial(\hA y)}
+ \frac{\partial(\delta^2\hat{\kappa}\sigma_{yz})}{\partial(\hA z)}
\notag\displaybreak[0]\\&=
\frac{\delta\hat\kappa}{\hat\ell}\frac{\partial\sigma_{xy}}{\partial x}
+ \frac{\hat\kappa}{\hA}\frac{\partial\sigma_{yy}}{\partial y}
+ \frac{\delta^2\hat\kappa}{\hA }\frac{\partial\sigma_{yz}}{\partial z}
\notag\displaybreak[0]\\&=
\frac{\hat\kappa}{\hA}\!\left(\!\delta^2\frac{\partial\sigma_{xy}}{\partial x}
+ \frac{\partial\sigma_{yy}}{\partial y}
+ \delta^2\frac{\partial\sigma_{yz}}{\partial z}\right),
\end{align}%
where $\delta = \hA/\hat\ell$ is used in the last line.  Doing this to the force balance in each direction gives
\begin{subequations}\label{eq:nondim_mom}\begin{align}
\pdr{\sigma_{xx}}{x} &+ \pdr{\sigma_{xy}}{y} + \pdr{\sigma_{xz}}{z} = 0,\label{eq:nondim_mom1}\\
\delta^2 \pdr{\sigma_{xy}}{x} &+  \pdr{\sigma_{yy}}{y} + \delta^2 \pdr{\sigma_{yz}}{z} = 0,\label{eq:nondim_mom2}\\
 \pdr{\sigma_{xz}}{x} & + \pdr{\sigma_{yz}}{y} + \pdr{\sigma_{zz}}{z}  = 0.\label{eq:nondim_mom3}
\end{align}\end{subequations}

We now observe the emergence of $\delta$ in the nondimensional equations~\eqref{eq:nondim_mom} arising from the scaling defined in~\eqref{eq:scales}.
This is what enables the asymptotic analysis that follows: for example, from~\eqref{eq:nondim_mom2} we see that $\partial\sigma_{yy}/\partial y = O(\delta^2)$, and so the through thickness variation in $\sigma_{yy}$ is small, and is quantified to be of size $O(\delta^2$).  This would not have been evident without nondimensionalizing.
Similarly, using the scaling in equation~\eqref{eq:scales} the dimensionless forms of the incompressibility and flow rule relations are
\begin{subequations}\begin{align}
\pdr{u}{x} + \pdr{v}{y} + \pdr{w}{ z}   &= 0 ,\label{eq:nondim_cont}\displaybreak[0]\\
\pdr{u}{x}&= \lambda s_{xx},\label{eq:nondim_flow_xx}\\
\pdr{v}{y}&= \lambda s_{yy},\label{eq:nondim_flow_yy}\\
\pdr{w}{z}&= \lambda s_{zz},\label{eq:nondim_flow_zz}\displaybreak[0]\\
\pdr{u}{y}+\delta^2 \pdr{v}{x}&= 2\delta^2 \lambda \sigma_{xy},\label{eq:nondim_flow_xy}\\
\pdr{u}{z}+\delta^2 \pdr{w}{x}&= 2\delta^2 \lambda \sigma_{xz},\label{eq:nondim_flow_xz}\\
\pdr{v}{z}+ \pdr{w}{y}&= 2\delta^2 \lambda \sigma_{yz}.\label{eq:nondim_flow_yz}
\end{align}\end{subequations}%
where the deviatoric stresses $s_{ij}$ is defined as 
\begin{align} \label{eq:hydrostatic}
{\sigma}_{ij} &= {s}_{ij}- {p}\delta_{ij} &
&\text{and}&
-{p}&=\frac{1}{3}({\sigma}_{xx}+{\sigma}_{yy}+\delta^2{\sigma}_{zz}).
\end{align}
From~\eqref{eq:hydrostatic} it can be seen that to get a non-zero pressure at leading order, $s_{zz}$ is also required to be the order of magnitude of $\hat \kappa$. 

With the scaling introduced, the von Mises yield criterion will be 
\begin{multline}
     (\sigma_{xx}-\sigma_{yy})^2+(\sigma_{xx}-\delta^2 \sigma_{zz})^2+(\sigma_{yy}-\delta^2 \sigma_{zz})^2
     \\
     + 6\delta^2\sigma^2_{xy}+ 6\delta^2\sigma^2_{xz}+ 6\delta^4\sigma^2_{yz}=6. \label{eq:nondim_yield}
\end{multline}
The Coulomb friction boundary condition on the roll surface $y=h(x)$ is $\vect{t\cdot\sigma\cdot n} = \mp\mu\vect{n\cdot\sigma\cdot n}$, where $\mu = \delta\beta$ is the coefficient of friction, $\vect{n} = (-\delta \intd h/\intd x, 1, 0)/\sqrt{1+\delta^2\big(\intd h/\intd x\big)^2}$ is the unit normal to the roll surface and $\vect{t} = (1, \delta \intd h/\intd x, 0)/\sqrt{1+\delta^2\big(\intd h/\intd x\big)^2}$ is the unit tangent in the rolling direction.  This may be expressed as
\begin{align}\label{eq:coulomb_this}
&\delta \frac{\intd h}{\intd x} \big(\sigma_{yy}-\sigma_{xx}\big) + \delta\!\left(\!1-\delta^2\!\left(\frac{\intd h}{\intd x} \right)^{\!\!2}\right)\! {\sigma}_{xy}
\\\notag&
= \mp \delta \beta \!\left(\!{\sigma}_{yy} - 2 \delta^2 \frac{\intd h}{\intd x}  {\sigma}_{xy} + \delta^2\!\left(\frac{\intd h}{\intd x}\right)^{\!\!2}\!{\sigma}_{xx}\!\right) \quad \text{on } y=h(x).
\end{align}
Throughout this work, the convention is adopted that the negative sign in $\mp$ corresponds to the region before the neutral point ($\hat{x} < \hat{x}_{\text{N}}$), while the positive sign applies to the region after the neutral point ($\hat{x} > \hat{x}_{\text{N}}$).

Assuming dimensional tensions $\hat F_{\mathrm{A/out}}$ are applied at location A and at the exit, the horizontal stress must satisfy
\begin{align} \label{eq:Fin/out}
    \frac{\hat F_{\mathrm{A/out}}}{4\hat \kappa \hA^2}
    &= \int_{-b_{\mathrm{A/out}}}^{b_{\mathrm{A/out}}} \int_{-h_{\mathrm{A/out}}}^{h_{\mathrm{A/out}}} \sigma_{xx} \, \intd y\intd z, & \quad \text{at} \, \text{A/exit}
\end{align}
where $h_{\text{out}}$ and $b_{\text{out}}$ are respectively half of the final thickness and width, and $\nhA =1$ and $\nbA$ are respectively half of the initial thickness and width at location A. 
Vertical symmetry requires the boundary conditions
\begin{align} \label{eq:sym}
    \sigma_{xy}(x,0,z) &= 0 &
    &\text{and}&
    v(x,0,z) &= 0.& 
\end{align}
There is no flow of material either through the roll surface $y=h(x)$, nor through the lateral edges of the wire $z=b(x)$. Mathematically, these are imposed as boundary conditions, expressed as
\begin{subequations}
\begin{align}
    \vect{V\cdot n} &= 0 &&\Rightarrow & v &= \frac{\intd h}{\intd x}u && \text{on} \, y= h(x),
\\
    \vect{V\cdot m} &= 0 &&\Rightarrow & w &= \frac{\intd b}{\intd x}u && \text{on} \, z= b(x),
\end{align}\label{eq:nopen}%
\end{subequations}%
where $\vect{V} = (u, \delta v, \delta w)$ is the velocity made dimensionless with $\UA$. Finally, integrating the incompressibility equation~\eqref{eq:nondim_cont} vertically from $y=-h(x)$ to $y=h(x)$, laterally from $z=-b(x)$ to $z=b(x)$, and axially from point A to any axial position $x$, using the divergence theorem and applying the boundary conditions~\eqref{eq:nopen} yields
\begin{equation} \label{eq:mass_ave}
  \int_{-b(x)}^{b(x)}  \int_{-h(x)}^{h(x)} u\ \mathrm{d} y  \mathrm{d} z= 4\nbA,
\end{equation}
where $h=u=1$ and $b=\nbA$ at point A gives the constant of integration as $4\nbA$.  Equation~\eqref{eq:mass_ave} effectively says that, since no material is lost or gained during rolling and the material is incompressible, the volume rate of flow across the wire cross-section is a constant independent of where along the wire it is measured at.

%%%%%%%%%%%%%%%%%%%%%%%%%%%%%%%%%%%%%%%%%%%%

\subsection{Leading-order asymptotic solution} \label{sec:solution}

Following non-dimensionalisation, all terms in the governing equations are now the same size except for the explicit factors of $\delta$, which is small.  Following the technique of asymptotics~\citep[see, e.g.][]{hinch}, we now expand in an asymptotic series in powers of the small parameter $\delta$, such that each successive term is asymptotically smaller than the preceding one.  Whilst many complicated expansions are sometimes needed, here a simple expansion in powers of $\delta$ is sufficient, as is the case for sheet rolling~\citep{erfanian2025}; this is suggested from the fact that the small parameter $\delta$ only appears as integer powers of $\delta$ in the governing equations and boundary conditions above, and may be verified a posteriori once the solution has been calculated.  We therefore expand any unknown variable as
\begin{equation}\label{eq:expansion}\phi= \phi^{(0)}+ \delta \phi^{(1)} + \delta^2 \phi^{(2)} + O(\delta ^3), 
\end{equation}
with $\phi^{(0)}$ not identically zero so that $\phi = O(1)$.  Here, $\phi$ represents any of the unknown variables $u$, $v$, $w$, $\sigma_{ij}$, $s_{ij}$, $p$ and $\lambda$; for example, $\sigma_{xx} = \sigma^{(0)}_{xx} + \delta\sigma^{(1)}_{xx} + \delta^2\sigma^{(2)}_{xx} + \cdots$. Substituting the asymptotic expansion into the governing equations~\eqref{eq:nondim_mom} to~\eqref{eq:mass_ave} allows smaller terms to be neglected relative to dominant ones successively, establishing an ordered hierarchy that leads to an approximate but systematically improvable solution with a quantifiable error. 
While this formalism could be used to compute the solution to an arbitrary order of accuracy (for example, \citet{erfanian2025} consider up to the second order terms $\sigma^{(2)}_{xx}$ for sheet rolling), here we will only consider the leading order terms.  
We therefore neglect all terms of $O(\delta)$ or higher to derive the leading-order equations. It should be noted that neglecting $O(\delta)$ terms in the expansion does not mean we are setting $\delta$ to zero, but rather 
that we are calculating the leading-order solution which represents the dominant physical behaviour for small, yet realistic, values of $\delta$, with a relative error of order $O(\delta)$ which could subsequently be corrected for by including higher-order terms.
From the hydrostatic pressure~\eqref{eq:hydrostatic} at leading order
\begin{equation} \label{eq:lead_p}
    \sigma^{(0)}_{yy} = -3p^{(0)} - \sigma^{(0)}_{xx}
\end{equation}
Substituting \eqref{eq:lead_p} into the yield function~\eqref{eq:nondim_yield} at leading order gives
\begin{subequations} \label{eq:lead_normal}\begin{align} 
\Big(\sigma^{(0)}_{xx} - (-3p^{(0)}&-\sigma^{(0)}_{xx})\Big)^2 \!+ \Big(\sigma^{(0)}_{xx}\Big)^2 \!+ \Big(3p^{(0)}+\sigma^{(0)}_{xx}\Big)^2\!=6\\\Rightarrow\qquad
    \sigma^{(0)}_{xx} &= -\frac{3p^{(0)}}{2} + \frac{\sqrt{4-3p^{{(0)} ^2}}}{2}, \\
    \sigma^{(0)}_{yy} &= -\frac{3p^{(0)}}{2} - \frac{\sqrt{4-3p^{{(0)} ^2}}}{2}.
\end{align} \end{subequations}

The force balance in the $y$ direction~\eqref{eq:nondim_mom2}, at leading order is reduced to
\begin{equation}
    \pdr{\sigma^{(0)}_{yy}}{y} = 0.
\end{equation}
Using this condition alongside the stress solutions~\eqref {eq:lead_normal} implies that $\sigma^{(0)}_{yy}$ and therefore $p^{(0)}$ and $\sigma^{(0)}_{xx}$ are vertically homogeneous which is the same as the slab method's assumption. We now show that these components are independent of $z$, as well, and only change along the rolling direction, $x$. From~\eqref{eq:nondim_flow_xy} and~\eqref{eq:nondim_flow_xz}, $u^{(0)}$ is independent of $y$ and $z$, which, together with the continuity equation~\eqref{eq:nondim_cont} and~\eqref{eq:nondim_flow_yz}, imply that $v^{(0)}$ is linear in $y$ and $w^{(0)}$ is linear in $z$. 
Now the set of equations~\eqref{eq:nondim_flow_xx}--\eqref{eq:nondim_flow_zz}, where their left-hand sides are only functions of $x$ while the stresses are functions of $p^{(0)}$ (equation~\eqref{eq:lead_normal}) shows that in order for a single value for $\lambda^{(0)}$ satisfies them all, $p^{(0)}$ must be a function of $x$ only. 

The force equilibrium in the $x$ direction~\eqref{eq:nondim_mom1} now can be integrated over $z$ and $y$ to give
\begin{equation}
    h(x)b(x) \sigma_{xx}^{(0)}(x) + b(x)\sigma_{xy}^{(0)} (x,h(x)) + h(x) \sigma_{xz}^{(0)} (x,b(x)) = 0,
\end{equation}
where $\sigma_{xy}^{(0)}(x,h(x))$ and $\sigma_{xy}^{(0)}(x,b(x))$ can be found from the boundary conditions~\eqref{eq:coulomb_this} and~\eqref{eq:edge_xz}, respectively. Therefore
\begin{equation} \label{eq:p_ode_b}
     hb \frac{\intd\sigma^{(0)}_{xx}}{\intd x} + \sigma^{(0)}_{xx} \!\left(\!b \frac{\intd h}{\intd x} + h \frac{\intd b}{\intd x} \right)\! +b \!\left(\!-\frac{\intd h}{\intd x}  \mp \beta \!\right)\! \sigma^{(0)}_{yy} = 0.
\end{equation}
This is the same equation as was derived using the slab method by~\citet{kazeminezhad2006pressure}, although our formulae for $\sigma^{(0)}_{xx}$ and $\sigma^{(0)}_{yy}$ here are different. 
Equation~\eqref{eq:p_ode_b} provides an ordinary differential equation (ODE) in terms of $p^{(0)}$ and $b(x)$, when combined with stress solutions~\eqref{eq:lead_normal}.

Another ODE results from the velocity equations.  Equations~\eqref{eq:nondim_flow_xy} and~\eqref{eq:nondim_flow_xz} at leading order show that $u^{(0)}$ has no $y$ or $z$ dependence and so is only a function of $x$. The volume flow rate~\eqref{eq:mass_ave} then requires that
\begin{equation} \label{eq:u_leading}
  u^{(0)}(x) = \frac{\nbA}{b(x)h(x)}.
\end{equation}
 From flow rules~\eqref{eq:nondim_flow_xx}--\eqref{eq:nondim_flow_zz}, $\lambda^{(0)}$ is a function of $x$. Therefore, from the tension flow rule in the $y$ direction~\eqref{eq:nondim_flow_yy} along with the symmetry condition~\eqref{eq:sym}, the vertical velocity is found as
\begin{equation} \label{eq:v_leading}
    v^{(0)} = y \lambda^{(0)} s^{(0)}_{yy}.
\end{equation}
Solutions~\eqref{eq:u_leading} and~\eqref{eq:v_leading}, along with the no-penetration surface condition~\eqref{eq:nopen}, results in
\begin{equation}\label{eq:lambda_v}
    \lambda^{(0)} = \frac{\nbA \intd h/ \intd x}{b h^2 s^{(0)}_{yy}},
\end{equation}
which holds not only on the surface but throughout the entire thickness, given that $\lambda^{(0)}$ is only a function of $x$.
Similarly, from the tension flow rule in the $z$ direction~\eqref{eq:nondim_flow_zz} coupled with no-flow normal to the edges condition~\eqref{eq:nopen},
\begin{equation}\label{eq:lambda_w}
    \lambda^{(0)} = \frac{\nbA \intd b/ \intd x}{b^2 h p^{(0)}}.
\end{equation}
Matching~\eqref{eq:lambda_v} and~\eqref{eq:lambda_w} gives an ODE for $b(x)$ as
\begin{equation} \label{eq:ode_b_vw}
  \frac { \intd b} {\intd x} = \frac{b}{h} \frac{\intd h} {\intd x} \frac{p^{(0)}}{s^{(0)}_{yy}}.
\end{equation}
By solving equations~\eqref{eq:ode_b_vw} together with~\eqref{eq:p_ode_b}, $b(x)$ and $p^{(0)}$ are solved. However, in equation~\eqref{eq:p_ode_b} the dependency on $b(x)$ can be removed to simplify further; 
an alternative ODE for $b(x)$ may be found by replacing $u^{(0)}$ from~\eqref{eq:u_leading} and $\lambda^{(0)}$ from~\eqref{eq:lambda_w} into flow rule~\eqref{eq:nondim_flow_xx}, as
\begin{equation}\label{eq:ode_b_u}
    b\frac{\intd h}{\intd x} + h \frac{\intd b}{\intd x} = - h \frac{\intd b}{\intd x} \frac{s^{(0)}_{xx}}{p^{(0)}}.
\end{equation}
Replacing expression~\eqref{eq:ode_b_u} into~\eqref{eq:p_ode_b} and further simplifying the resultant using~\eqref{eq:ode_b_vw} gives
\begin{equation}\label{eq:stress_ode}
         \frac{\intd\sigma^{(0)}_{xx}}{\intd x} -  \!\left(\! \frac{\intd h/ \intd x}{h} \frac{s^{(0)}_{xx}}{s^{(0)}_{yy}} \right)\! \sigma^{(0)}_{xx} +  \frac{1}{h}\! \left(\!-\frac{\intd h}{\intd x}  \mp  \beta\! \right) \!\sigma^{(0)}_{yy}  = 0,
\end{equation}
which is an ODE for the stresses without depending on the width. From~\eqref{eq:Fin/out}, the boundary conditions at leading order at point A and the roll-gap exit are
\begin{subequations} \label{eq:Fin/out_lead}
\begin{align}
    4\nbA\sigma^{(0)}_{xx}(x=\mathrm{A}) &= F_{\mathrm{A}} && \text{at point A,} \\
    4bh\sigma^{(0)}_{xx}(x=1) &= F_{\mathrm{out}} && \text{at exit.}
\end{align}
\end{subequations}
With this, equation~\eqref{eq:stress_ode} can be solved from point A forward with $-$ve sign and from the exit backwards with $+$ sign.
It is interesting to note from equations~\eqref{eq:ode_b_vw} and~\eqref{eq:p_ode_b} that $b(x)$, and consequently the width of contact, $W_c$, and the lateral spread of the wire, $W_t$, depend only on the initial diameter of the wire, reduction ratio, roll radius, and friction coefficient, and not on the roll speed on the material type (e.g.\ the yield stress $\hat\kappa$).  The influence of roll speed and material type would become apparent if hardening were incorporated into the analysis, although they likely remain negligible in comparison.
It is worth commenting on the order of magnitude of the error. Based on the scaling in~\eqref{eq:scales} and asymptotic expansion~\eqref{eq:expansion},
\begin{equation}
    \hat b = \hA b = \hA \big(b^{(0)} + \delta b^{(1)} + O(\delta^2) \big),
\end{equation}
the solution obtained up to the leading order, $b^{(0)}$, carries a relative error of order $O(\delta)$.

%%%%%%%%%%%%%%%%%%%%%%%%%%%%%%%%%%%%%%%%%%%%%%%%%%%

\subsection{Summary and numerical evaluation}\label{sec:numerics}

In this section, a simple numerical procedure for performing the calculations is detailed.

Equation~\eqref{eq:stress_ode} can be written in terms of $p^{(0)}$ by using
\begin{equation} \label{eq:chainru}
    \frac{\intd \sigma_{xx}^{(0)}}{\intd x} = \frac{\intd \sigma_{xx}^{(0)}}{\intd p^{(0)}} \frac{\intd p^{(0)}}{\intd x}, 
\end{equation}
and replacing stress components from~\eqref{eq:lead_normal} and~\eqref{eq:hydrostatic}.
As described by equation~\eqref{eq:Fin/out_lead}, the force at point A is necessary for solving the ODE~\eqref{eq:stress_ode}.  However, it is assumed that force (and as a result pressure) at point A is the same as that at the roll-gap entrance. This assumption will be justified in the following section using FE data, which shows a pressure drop following the large peak at the entrance.
The exit point in the model is the same as the roll-gap exit, therefore, $F_{\mathrm{out}}$ determines the pressure at the exit. If there is no exit tension, then $\sigma_{xx}^{(0)}$ becomes zero at the exit. Otherwise, the value of $\sigma_{xx}^{(0)}$ at the exit depends on the final width. In such a case, $b$ at the exit must be estimated and iteratively refined to align with the width determined from~\eqref{eq:ode_b_vw} or~\eqref{eq:ode_b_u}.

The solution for $p^{(0)}$ is chosen to satisfy the forward and backward tension conditions, which are taken to be zero for the results presented below.
Therefore,  $p^{(0)}$ is solved by integrating equation~\eqref{eq:stress_ode} forward from the entrance with positive sign of friction, and integrating equation~\eqref{eq:stress_ode} backwards from the exit with positive sign of friction, using the \Matlab\ ODE solver \texttt{ode45} \citep{MATLAB:R2024a_u1}. 
This is the same solution as the slab method, and the two curves thus produced are referred to as the pressure hill, and the intersection determines the location of the neutral point.

After solving for stresses, $b(x)$ and consequently the lateral spread, $W_t$, is determined by integrating either equation \eqref{eq:ode_b_vw} or \eqref{eq:ode_b_u} from point A to the exit. The integration is performed using the \Matlab\ ODE solver \texttt{ode45}, with the initial condition $b(0)=\nbA$.
For the rolling of a cylindrical wire, we assume $\nbA = \nhA = 1$, as stated in section~\ref{sec:cylindrical} and further justified in \ref{app:rectangular}, and consistent with the non-dimensionalisation. Nevertheless, retaining $\nbA$ explicitly in the governing equations allows the same model to be applied to subsequent roll stands, where the initial cross-section is no longer circular but approximately rectangular with bulged edges. In such cases, location $A$ corresponds to the roll-gap entrance.
As detailed in the introduction and shown in figure~\ref{fig:widthschematics}, the area of the rectangular cross-section is assumed to be equal to that of the actual wire in the roll gap. As a result, the lateral spread is determined by equating the cross-sectional area of the real shape to that of the rectangle, with $\hat W_c$ the width in contact with the roll and $\hat W_t$ the total lateral spread then given by~\eqref{eq:w_contact} and~\eqref{eq:w} respectively.

Recall that $\hat{b}$ in equations~\eqref{eq:w_contact} and~\eqref{eq:w} is defined as $\hat{b} = \hA b$, where $b$ is determined independently of the material type. As a result, both $W_c$ and $W_t$ do not depend on the wire material.
The total computation time varies depending on the tension at the exit, and is at most in the order of seconds on a standard laptop. For zero tension, solving stresses and lateral spread takes $0.1$ seconds on a single core of a 2019 Intel i7-8665U laptop CPU. 

%%%%%%%%%%%%%%%%%%%%%%%%%%%%%%%%%%%%%%%%%%%%%%%%%%

\section{Experimental Methodology}\label{sec:exp}
Experimental verification was carried out using a Hille 25 rolling mill using D2 tool steel rollers with a roll diameter of $100\,\mathrm{mm}$. The mill operates at a typical speed of $60\,\mathrm{rpm}$.
Stainless steel 316 rods of initial diameters 2.96, 3.96, 5.96 and $7.96\,\mathrm{mm}$ were rolled without lubrication at room temperature to reductions of 30--60$\%$ in a single pass. 
The initial and final lengths, thicknesses and widths were all measured using a Mitutoyo 293-240-30 micrometer.

%%%%%%%%%%%%%%%%%%%%%%%%%%%%%%%%%%%%%%%%%%%%%%%%

\section{Results and discussion} \label{sec:results}

\subsection{Lateral spread} \label{sec:results_width}

\begin{figure*}[t]
    \centering
    \includegraphics[width=1\linewidth]{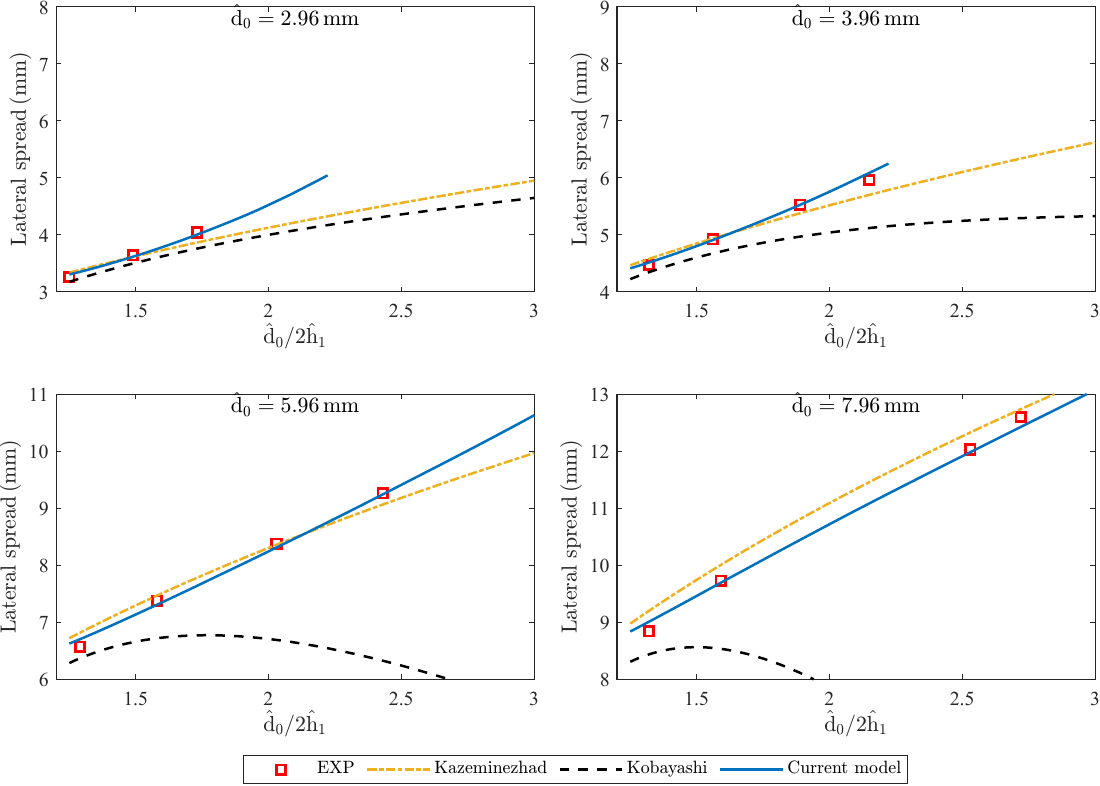}
    \caption{Lateral spread for stainless steel wires with different diameters, $\hat d_0$, and reduction ratios, $\hat d_0/(2\hat h_1)$ from different methods; experimental data, empirical equation~\ref{eq:kaztotalwidth2} from~\citet{kazeminezhad2005width}, empirical equation from Kobayashi~\eqref{eq:kobayashi}, and the current model. Other parameters used are $(\hat R, \mu)=( 50\, \mathrm{mm}, 0.25)$.}
    \label{fig:lateral spread}
\end{figure*}

The lateral spread prediction, $\hat W_t$, for both cases, is plotted in Figure~\ref{fig:lateral spread} for various diameters and reduction ratios with 50~$\mathrm{mm}$ roll radius.
The results are compared with experimental data, Kobayashi's empirical equation~\eqref{eq:kobayashi}, and empirical equation~\eqref{eq:kaztotalwidth2} due to~\citet{kazeminezhad2005width}. The latter equation is particularly relevant because the parameters in~\eqref{eq:kaztotalwidth2} were fitted for stainless steel. 
In the current model, predictions depend on the friction coefficient, which is challenging to determine experimentally. The experimental data were obtained under non-lubricated test conditions, and a friction coefficient of $\mu=0.25$ has been shown to provide the best agreement with experimental data, consistent with the range typically reported for flat rolling of metal wires~\citep{lin2008finite,lee2018development}. Consequently, this value is used for the results in Figure~\ref{fig:lateral spread}. 
 The current model fails at larger reductions for $\hat d_0=2.96\ \mathrm{mm}$ and $\hat d_0=3.96\ \mathrm{mm}$, as indicated by the unfinished lines, due to a singularity in pressure (see equation~\ref{eq:lead_normal}); interestingly, the experiments also encountered difficulties achieving comparable larger reductions during single-pass rolling in these cases, and so experimental data in these cases is also lacking.  Where data is available, the model agrees closely with the experimental data for all diameters and reduction ratios.  Kazeminezhad Equation~\eqref{eq:kaztotalwidth2} depends only on the reduction ratio. While it performs well for smaller reduction ratios, it deviates for larger values, with the deviation appearing to depend on wire diameter (e.g., underestimates for $\hat d_0 = 5.96\ \mathrm{mm}$ and overestimates for $\hat d_0 = 7.96\ \mathrm{mm}$). Kobayashi's equation depends on roll radius and wire diameter as well as the reduction ratio and seems to be derived for small wire diameter conditions. It underestimates the lateral spread across all wire diameters, particularly at higher reduction ratios, and is the least reliable for predicting the set of experimental data presented in Figure~\ref{fig:lateral spread}.

\begin{figure*}[t]
    \centering
    \includegraphics[width=1\linewidth]{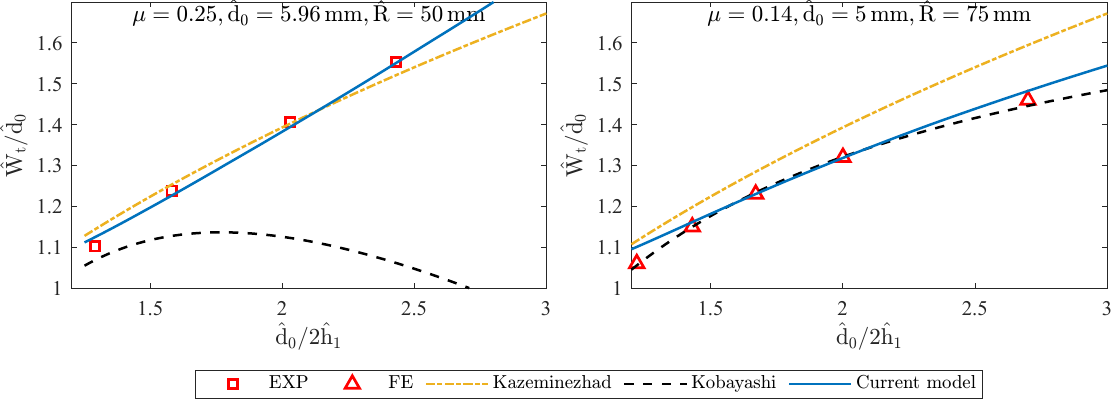}
    \caption{Effect of friction coefficient on lateral spread. Results are plotted from different methods; experimental data, FE simulations~\citep{vallellano2008analysis}, empirical equation~\ref{eq:kaztotalwidth2} from~\citet{kazeminezhad2005width}, empirical equation from Kobayashi~\eqref{eq:kobayashi}, and the current model.}
    \label{fig:mu_effect}
\end{figure*}

  Some studies suggest that the friction coefficient has a negligible effect on lateral spread~\citep{carlsson1998contact,kazeminezhad2008error}, attributing this to the movement of lubricant toward the roll edges, driven by the extremely high contact pressure at the entry point.
   Similarly, the empirical equations~\eqref{eq:kaztotalwidth2} and~\eqref{eq:kobayashi} exhibit no dependence on friction.
  To examine the impact of friction, the ratio of lateral spread to the initial wire diameter is plotted in Figure~\ref{fig:mu_effect} for two different friction coefficients.
 In the absence of experimental data under the lubricated condition, the FE results from~\citet{vallellano2008analysis} are utilised. 
The data reported by~\citet{vallellano2008analysis} corresponds to a wire with a 5~$\mathrm{mm}$ diameter, rolled using 75~$\mathrm{mm}$ radius rolls, with no applied forward or backward tension. 
The yield stress, $\hat Y$, is assumed to be constant and equal to 385~$\mathrm{MPa}$ the same as the average value in FE simulations. \Citet{vallellano2008analysis} used the Tresca friction law, $\hat \tau = m \hat \kappa$ with the coefficient of $m=0.25$ to generate their FE results.
Generally, the relationship $\mu< m/\sqrt{3}$ is used essentially for a simple elastoplastic model~\citep{zhang2016relationship}, where the upper limit, $\mu=0.14$, is used to transfer $m$ into the Coulomb friction coefficient, $\mu$.
The data represented by the black line with circles corresponds to the experimental measurements for $\hat d_0=5.96~\mathrm{mm}$ without lubrication ($\mu=0.25$), while the counterpart for $\hat d_0=5~\mathrm{mm}$ is derived from FE simulations using $m=0.25$ (approximated by $\mu=0.14$). 
To minimise the influence of different wire diameters in this comparison, wires with relatively similar diameters were selected, and the results were scaled by the initial diameter.

The comparison in Figure~\ref{fig:mu_effect}
indicates that friction does influence lateral spread, particularly at higher reductions. Greater friction appears to increase lateral spread. This can be explained by the increased constraint in the longitudinal direction. Although friction acts in both the lateral and rolling directions, the lateral contact length is smaller than the longitudinal contact length. As a result, friction primarily acts as a resistance in the rolling direction.
This effect is accurately captured by the current model. For this wire diameter, Kazeminezhad's equation correctly predicts lateral spread for larger friction coefficients but tends to overestimate the results for lower friction values. Conversely, Kobayashi's equation demonstrates better agreement under low-friction conditions compared to high friction, although different roll and wire diameters might also influence this.
%%%%%%%%%%%%%%%%%%%%%%%%%%%%%%%%%%%%%%%%%%%%%%%%

\subsection{Longitudinal spread} \label{sec:length}

The model can be easily used to predict the final length of the wire.
Since the volume of metal in the wire remains constant, the wire length varies inversely proportional to the cross-sectional area.
The final length of the flattened write, $\hat{L}_1$ can therefore be calculated from
\begin{equation}
 \pi\frac{{\hat d_0}^2}{4}\,\hat{L}_0  =  4 \hat b_1 \, \hat h_1 \, \hat{L}_1,
\end{equation}
where $\hat{L}_0$ is the initial wire length, $\hat h_1$ is the final half thickness and $\hat b_1$ the final half width from equation~\eqref{eq:ode_b_vw}.

\begin{figure*}[t]
    \centering
    \includegraphics[width=1\linewidth]{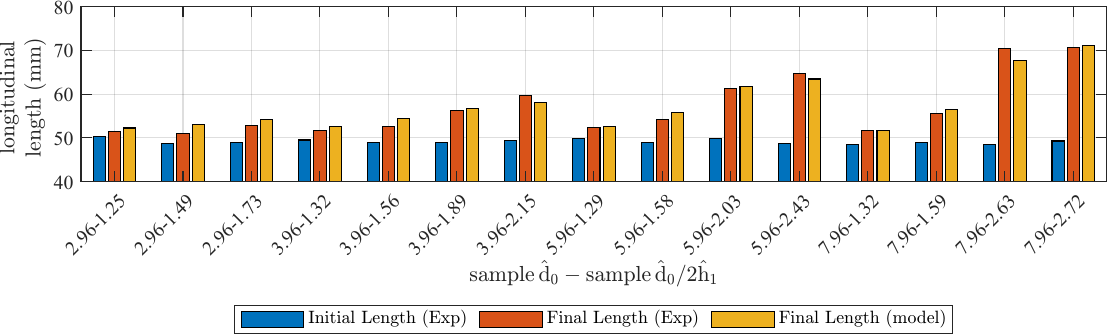}
    \caption{The initial wire length, along with the final experimental and predicted length for different samples. Each sample is labelled according to its initial wire diameter and the ratio of initial diameter to final thickness (for instance, ``2.96--1.25'' refers to a wire with an initial diameter of 2.96~mm and a reduction ratio of 1.25). Other parameters used are $(\hat R, \mu)=( 50\, \mathrm{mm}, 0.25)$.}
    \label{fig:length_error}
\end{figure*}%
Figure~\ref{fig:length_error} shows the initial length along with the final experimental and predicted length for different samples.
Each sample is labelled according to its initial wire diameter and the ratio of initial diameter to final thickness (for instance, ``2.96--1.25'' refers to a wire with an initial diameter of 2.96~mm and a reduction ratio of 1.25).
Overall, the predicted final lengths show good agreement with the experimental measurements across the tested range of samples: the average absolute error in length prediction is $1.1\,\mathrm{mm}$, and the average relative error in length prediction is $2\,\%$.
This confirm that the model reasonably captures the longitudinal deformation associated with lateral spread and reduction during wire rolling.
%%%%%%%%%%%%%%%%%%%%%%%%%%%%%%%%%%%%%%%%%%%%%%%%

\subsection{Roll pressure} \label{sec:results_press}
\begin{figure*}[t]
    \centering
    \includegraphics[width=1\linewidth]{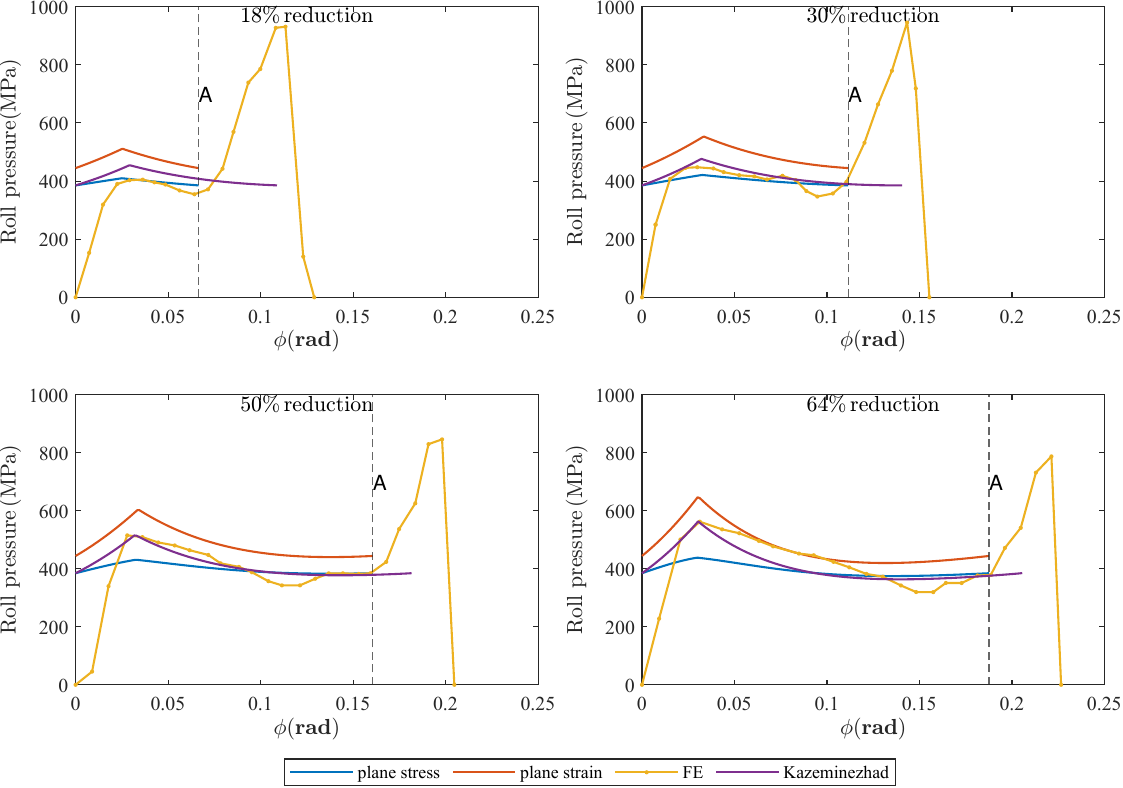}
    \caption{Roll pressure from different methods; plane-strain slab method~\citep{erfanian2025}, current plane-stress method, and FE simulations~\citep{vallellano2008analysis} and \citet{kazeminezhad2006pressure} slab method. Plotted for different reduction ratios against contact angle $\phi$. Other parameters used are $(\hat \kappa,\hat d_0, \hat R, \mu)=(385/\sqrt{3}\, \mathrm{MPa},5\,\mathrm{mm}, 75\,\mathrm{mm}, 0.14)$.}
    \label{fig:sigma_yy}
    \end{figure*}

 The roll pressure $\sigma^{(0)}_{yy}$ is found from~\eqref{eq:lead_normal} and plotted with respect to the contact angle, $\phi$ in Figure~\ref{fig:sigma_yy} for different reduction ratios.
 Results are compared with FE simulations from~\citet{vallellano2008analysis}, the slab model from~\citet{kazeminezhad2006pressure} and the plane-strain slab model for the rolling of thin sheet in~\citet{erfanian2025}. The same parameters as those used in~\citet{vallellano2008analysis} and described in the previous section are applied.
 The results for plane stress and plane strain are calculated from point A in Figure~\ref{fig:widthschematics}.
 
\begin{figure*}
    \centering
    \includegraphics[width=1\linewidth,height=0.24\textheight,keepaspectratio]{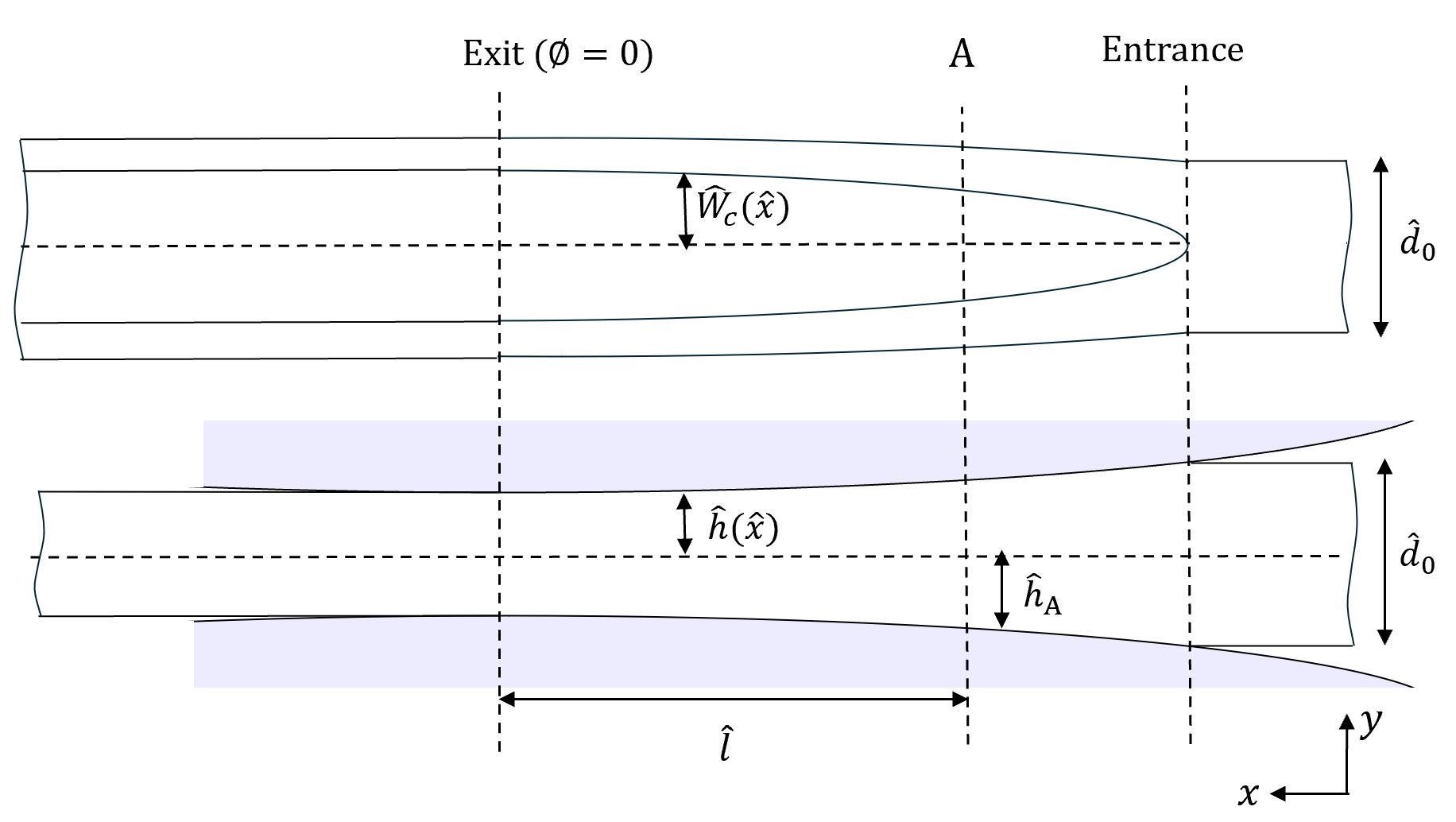}
    \caption{A diagram of the deforming wire, seen from above (top figure) and from side (bottom figure). For symbols, refer to the main text. Adapted from Figure~3 of~\citet{carlsson1998contact}.}
    \label{fig:topview}
\end{figure*}

FE results for all reductions can be seen to have two distinct regions; a massive rise between the entrance and point A and a shallow rise from point A to the exit. This trend can be explained better when looking alongside Figure~\ref{fig:topview}
which schematically shows how the wire looks as seen from the side and above while rolling. The wire enters between rolls from the right-hand side with a circular cross-section and exits from the left with an almost rectangle cross-section, therefore, the contact surface forms rather like half of an ellipse. 
As explained by \citet{carlsson1998contact} when the material starts to deform, surrounding parts of the wire are still in the elastic range and will therefore resist deformation. The situation may be compared to that of an indentation. As a result, a normal pressure will build up, resulting in a sharp rise in contact pressure soon after the entrance. As deformation continues, larger parts of the wire deform, reducing resistance from the remaining elastic regions, which in turn results in a pressure drop until point A. At this point, the material begins to flow laterally, and a typical friction hill develops, with a pressure peak forming between this point and the roll exit~\citep{carlsson1998contact}.  

From the results, it can be seen that the current model correctly predicts the location of point A by assuming that the area of the cross-section at this point is the same as that of the wire before rolling. Both the current plane-stress model and~\citet{kazeminezhad2006pressure} correctly predict the roll pressure at point A, for larger reductions. From point A to the roll exit, the FE results lie between the plane-strain and plane-stress predictions, reflecting the presence of 3D material flow within the roll gap. Consequently, the results presented by~\citet{kazeminezhad2006pressure} show better agreement with the FE data.
Yet, despite the model underestimates the pressure, it consistently provides accurate predictions of lateral spread. This may be attributed to the importance of the stress ratio $s_{zz}/s_{yy}$ --- rather than the absolute magnitudes --- in governing the lateral deformation of the wire. The model appears to capture this ratio correctly, despite discrepancies in individual stress components.
Another notable difference between the FE results and the perfect plastic models in Figure~\ref{fig:sigma_yy} is that the roll pressure starts at zero at the entrance and returns to zero at the exit. This behaviour arises from the inclusion of elasticity in the FE simulations, which is absent in the perfect plastic models.

%%%%%%%%%%%%%%%%%%%%%%%%%%%%%%%%%%%%%%%%%%%%%%%%

\section{Parametric study} \label{sec:parametric_study}

The present model, and indeed the governing equations behind it, depend only on three key non-dimensional parameters: the friction coefficient $\mu$, the aspect ratio $\delta$, and the reduction ratio $2\hat{h}_1 / \hat{d}_0$ (recall $\hat{h}_1$ is the final half-thickness and $\hat{d}_0$ is the initial wire diameter).  
The aspect ratio $\delta = \hA/\hat\ell$ characterises the geometry of the deformation zone, and can also be expressed using~\eqref{eq:h0} and~\eqref{equ:ell-dimensional} as
\begin{equation}\label{equ:ell-dimensionless}
\delta = \frac{\hA}{\hat \ell} = \frac{\sqrt{\pi}}{4}\!\!\left(\!\frac{\hat R}{\hat d_0}\!\!\left(\!\!\frac{\sqrt{\pi}}{2}- \frac{2\hat h_1}{\hat d_0}\!\right)\! - \frac{1}{4}\!\!\left(\!\!\frac{\sqrt{\pi}}{2}- \frac{2\hat h_1}{\hat d_0}\!\right)^{\!\!2}\,\right)^{\!\!\!-1/2}\!\!\!\!.
\end{equation}
Here, $\hat R/\hat{d}_0$ is also a non-dimensional parameter, and is arguably more intuitively connected to the dimensional rolling problem than is $\delta$.  In what follows, therefore, we perform a parametric study by varying the three independent dimensionless parameters $\mu$, $2\hat{h}_1/\hat{d}_0$, and $\hat R/\hat{d}_0$, with $\delta$ then calculated using~\eqref{equ:ell-dimensionless}.

%%%%%%%%%%%%%%%%%%%%%%%%%%%%%%%%%%%%%%%%%%%%%%%%

\subsection{Range of validity} \label{sec:range_validity}

\begin{figure*}[t]
    \centering
    \includegraphics[width=1\linewidth]{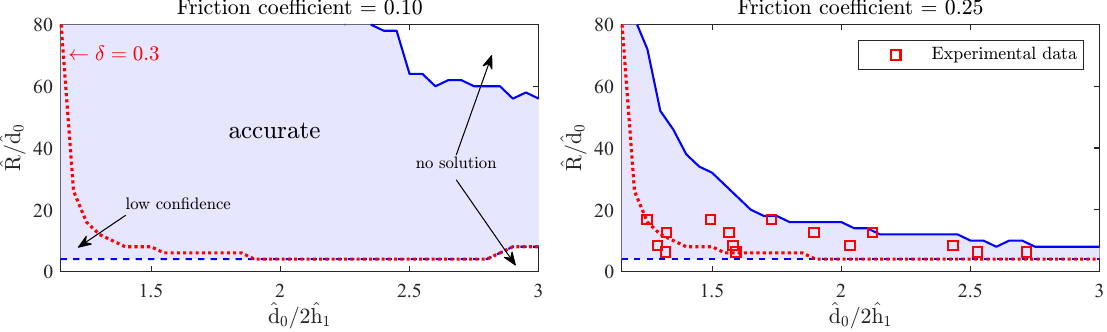}
    \caption{The range of validity of model at different reduction ratios and $\hat{R}/\hat{d}_0$ for $\mu=0.1$ (left panel) and for $\mu=0.25$ (right panel). The shaded region represents the operating range of the model for which the normal stresses (equation~\eqref{eq:lead_normal}) are non-imaginary values. Below the red dotted line related to $\delta > 0.3$ and the prediction should be interpreted with caution. The experimental data in the right panel are those reported in Figure~\ref{fig:lateral spread}.}
    \label{fig:valid_range}
\end{figure*}

The range of validity of model  at different reduction ratios and $\hat{R}/\hat{d}_0$ is shown for two friction coefficients in Figure~\ref{fig:valid_range}.
For each combination of parameters, the valid range is determined by ensuring the computed normal stresses (equation~\eqref{eq:lead_normal}) remain real quantities (i.e., the expression under the square root remains positive). 
A more restrictive criterion for defining the range of validity of the model is to exclude large values of $\delta$, as dictated by the asymptotic formulation governing the solution accuracy.
By way of indication, the line corresponding to $\delta=0.3$ is plotted in figure~\ref{fig:valid_range}, with the region below this curve corresponding to $\delta > 0.3$.
Therefore, when the model is used to predict the width in cases where the roll and wire sizes are comparable and the reduction ratio is small, the results should be interpreted with caution, although some experimental data points presented in Figure~\ref{fig:lateral spread} are located in $\delta > 0.3$ region and the comparison in Figure~\ref{fig:lateral spread} already shows good accuracy.

For a small friction coefficient, the model remains valid over a wider range of geometric parameters. In contrast, increasing the friction coefficient to $\mu = 0.25$ considerably narrows the range of validity, as higher friction amplifies the $p^{(0)}$ term in equation~\eqref{eq:lead_normal}.

Overall, the results show that the asymptotic model is most reliable for small aspect ratios (large $\hat{R}/\hat{d}_0$) and moderate reductions, which are representative of practical wire-rolling conditions. 
This is also consistent with the assumption of small $\delta$ in the asymptotic solution.
It is worth noting that exceeding the max or min lines not only invalidates the present model (due to the square root going negative), but also correlates to a lack of obtainable result, as discussed in section~\ref{sec:results_width}; it may well be, therefore, that the physical rolling process undergoes a qualitative change beyond these parameters, and so these minima and maxima lines are dictated not by the approximations of the model but by the physics of the rolling process.

%%%%%%%%%%%%%%%%%%%%%%%%%%%%%%%%%%%%%%%%%%%%%%%%

\subsection{The effects of parameters on lateral spread}\label{sec:parameters_effect}

\begin{figure*}[t]
    \centering
    \includegraphics[width=\linewidth]{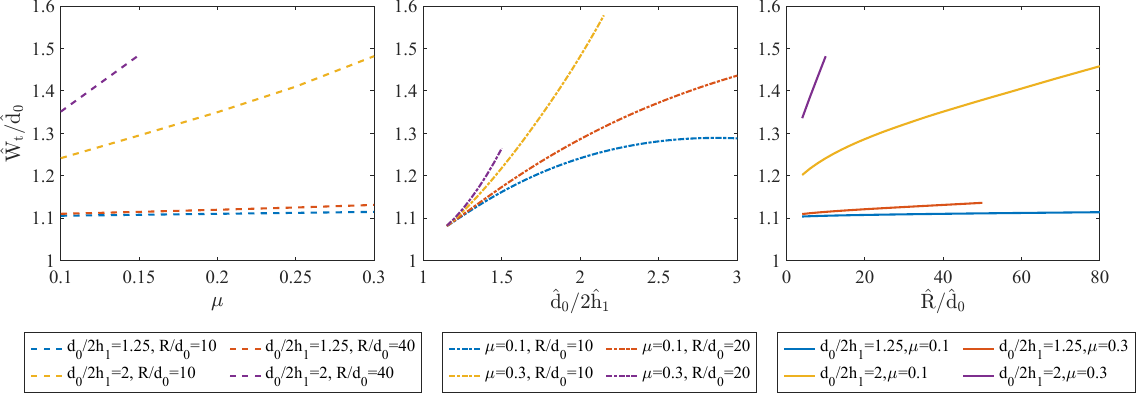}
    \caption{The effect of friction $\mu$ (left panel), reduction ratio $\hat{d}_0/2\hat h_1$ (middle panel) and roll size $\hat R/\hat d_0$ (right panel) on the predicted relative lateral spread $\hat{W}_t/\hat{d}_0$.}
    \label{fig:parametric_study}
\end{figure*}%
Figure~\ref{fig:parametric_study} summarises a parametric study of the effect of different non-dimensional parameters on predicted lateral spread over the range of validity shown in Figure~\ref{fig:valid_range}.
The effect of the friction coefficient is shown in the left panel. 
Over the examined range $\mu\in[0.1,0.3]$, the width is only weakly sensitive to changes in friction for the low-reduction cases $\hat{d}_0/2\hat{h}_1=1.25$, for any roll sizes (blue and brown dashed lines). 
For the larger reduction case $\hat{d}_0/2\hat{h}_1=2$ the effect of friction is more pronounced (yellow and purple dashed lines): higher $\mu$ produces a noticeably larger lateral spread. Physically, this behaviour reflects that, at small reductions, material flow is only weakly controlled by tangential tractions, whereas at larger reductions friction increasingly impedes longitudinal exit flow and promotes lateral displacement, so $\mu$ becomes more influential.

As expected, all curves show a clear monotonic increase of lateral spread with increasing reduction ratio in the middle panel of Figure~\ref{fig:parametric_study}.
However, the rate of increase is controlled by both $\mu$ and $\hat{R}/\hat{d}_0$. 
For the low-friction case ($\mu=0.10$, blue and brown curves), the width increases steadily but tends to flatten for large reductions, while for the higher-friction case ($\mu=0.30$, yellow and purple lines) the width rises much more rapidly with reduction. 
This demonstrates that friction amplifies the effect of reduction on lateral spread.

Finally, on the right panel of Figure~\ref{fig:parametric_study}, the influence of roll size $\hat{R}/\hat{d}_0$ depends on the reduction and friction. 
For the mild reduction case ($\hat{d}_0/2\hat{h}_1=1.25$, blue and brown lines), the relative width is almost insensitive to $\hat{R}/\hat{d}_0$ over the plotted range. 
By contrast, for the larger reduction ($\hat{d}_0/2\hat{h}_1=2$, yellow and purple lines), the width increases noticeably with $\hat{R}/\hat{d}_0$, and the increase is larger for higher friction ($\mu=0.3$, purple) than for low friction ($\mu=0.1$, yellow). 
A possible interpretation is that increasing $\hat{R}/\hat{d}_0$ increases the effective contact length and reduces local curvature effects, allowing more longitudinal material displacement to be converted into lateral spread for the higher-reduction cases; the conversion is enhanced when friction is larger.

%%%%%%%%%%%%%%%%%%%%%%%%%%%%%%%%%%%%%%%%%%%%%%%%

\section{Conclusion} \label{sec:conclusion}

\begin{figure*}[t]
\includegraphics[width=1\linewidth]{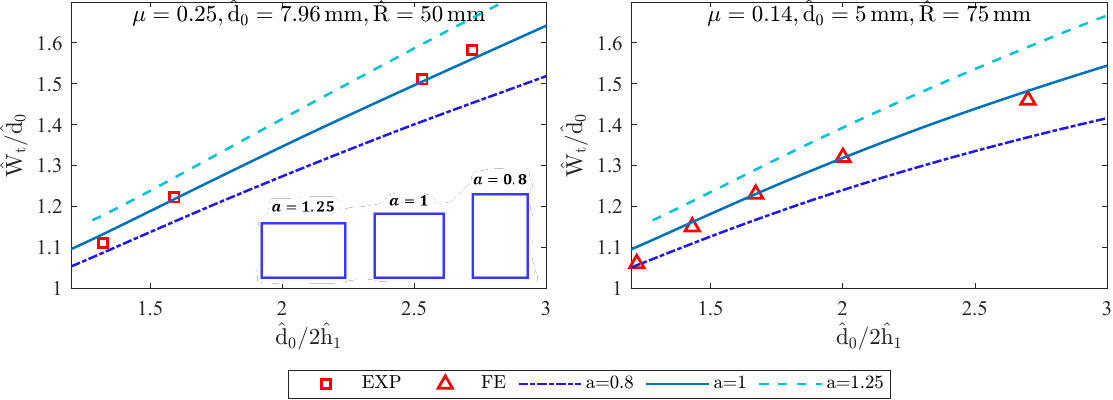}
\caption{Effect of the assumed cross-sectional shape at location~A on the predicted lateral spread. 
Different shapes corresponding to different values of \(a = \bA / \hA\) are illustrated in the left panel.}
\label{fig:a}
\end{figure*}%

This work develops a mathematical framework for wire rolling, emphasising the accurate prediction of lateral spread. A general state of deformation is simplified to the plane-stress deformation with the assumption of a small width with respect to the length. The set of assumptions made leads to the simplest possible first-principles model, which is shown to successfully predict lateral spread across a range of wire diameters, reduction ratios, and different friction coefficients. Importantly, this model achieves accurate predictions without requiring empirical fitting parameters or assuming additional factors. 
The validity of the model is demonstrated by comparison with both experimental and FE results, and the range of validity is defined. Interestingly, the range of validity also correlates with a lack of obtainable experimental result, suggesting an implication for the physics of the rolling process.

The assumption of plane-stress deformation is not strictly satisfied; comparison of roll pressure predictions with FE results suggests that the actual pressure distribution falls between that of plane-stress and plane-strain deformation. A possible explanation for this lies in the importance of the stress ratio, $s_{zz}/s_{yy}$, rather than the absolute stress values, in predicting the lateral spread; while the model may underestimate individual stresses, it appears to capture the correct ratio, leading to accurate predictions of lateral spread.

Nonetheless, having formulated stress and strain equations in all directions lays the foundation for a more general model. 3D numerical studies conducted using \Abaqus\ software by~\citet{carlsson1998contact} and~\citet{vallellano2008analysis} show that the contact pressure distribution is more complex than the traditional pressure hill profile. Relaxing the assumption of a small rate of change in width in the current model would allow the problem to be modelled under a general stress state. Also, solving the higher order terms would enable accounting for the curvature of the edges.

Cold-rolled wire often exhibits a strong texture in the axial direction, while the texture in the cross-section can be quite different~\citep{carlsson1998contact}.
The von Mises function used to describe the yielding condition could be replaced with an anisotropic yield criterion (e.g., Hill’s 1948 yield criterion) with known anisotropy coefficients. 
In such a case, the solution structure would largely remain unchanged, except for the incorporation of weighted stresses. 
However, this scenario is of limited scientific significance because it is the rolling process itself which induces anisotropy. 
To accurately capture the anisotropy effect, it would be necessary for the anisotropy to evolve throughout the process.
This would require a model describing how deformation induces anisotropy within the material, thereby adding an additional layer of complexity to the modelling process and is currently work in progress.
Additionally, further studies are required to explore potential correlations between the challenges faced in experiments when achieving certain reductions in a single pass and the pressure singularity predicted by the current model.

%%%%%%%%%%%%%%%%%%%%%%%%%%%%%%%%%%%%%%%%%%%%%%%%

\section*{Acknowledgements}
The authors are grateful to Dr~Lander Galdos (Mondragon University) for initial discussions concerning the lack of models for wire flat rolling.
ME gratefully acknowledges the support of a University of Warwick Chancellor's Scholarship.
EJB is grateful for the UKRI Future Leaders’ Fellowship funding (MR/V02261X/1) supporting this work.
For the purpose of open access, the authors have applied a Creative Commons Attribution (CC BY) license to any Author Accepted Manuscript version arising from this submission.

\appendix
\section{Effect of initial cross-sectional shape} \label{app:rectangular}

All results here have assumed that when a cylindrical wire is rolled, it first transforms into a square wire with aspect ratio $\bA/\hA = a = 1$, as this preserves the symmetry of the cylindrical wire.  In this appendix, we further investigate the effects of this assumption.
Figure~\ref{fig:a}, computed as described in section~\ref{sec:numerics} but for $\nbA=a$, shows similar results to figure~\ref{fig:mu_effect} but details the effect of varying $a$ on the results of our model.   Figure~\ref{fig:a} justifies that the choice of $a=1$ not only preserves symmetry but also gives the most accurate results compared with experimental results.  Similar results were obtained for all experimental wire diameters, although Figure~\ref{fig:a} plots the experimental results for the $\hat d_0 = 7.96\,\mathrm{mm}$ wire as these showed the greatest variation with $a$.

\raggedright
\bibliography{references}

@string(JMPT="J. Mater. Process. Technol.")

@string(IJMS="Int. J.~Mech. Sci.")

@misc{Hill1955,
  author = {R. Hill},
  title = {Private communication to {B.I.S.R.A.}},
  year = {1955},
  month = {November},
  day = {16},
  note = {letter to A. W. McCrum},
}

@book{hinch,
    author="E. J. Hinch",
    year="1991",
    title="Perturbation Methods",
    publisher="Cambridge",
    doi = {10.1017/CBO9781139172189}
}

@article{erfanian2025,
      title={Through-Thickness Modelling of Metal Rolling using Multiple-Scale Asymptotics}, 
      author={Mozhdeh Erfanian and Edward James Brambley and Francis Flanagan and Doireann O'Kiely and Alison N. O'Connor},
      year={2025},
      journal = {Eur.~J.~Mech.~A/Solids.},
      volume = {113},
      pages = {105712},
      doi = {10.1016/j.euromechsol.2025.105712},
}

@article{flanagan2025,
      title={Careful Finite Element Simulations of Cold Rolling with Accurate Through-Thickness Resolution and Prediction of Residual Stress}, 
      author={Francis Flanagan and Alison N. O'Connor and Mozhdeh Erfanian and Omer Music and Edward James Brambley and Doireann O'Kiely},
      year={2025},
      journal={Eur.~J.~Mech.~A/Solids.},
      volume={114},
      pages={105761},
      doi={10.1016/j.euromechsol.2025.105761},
}

@article{Kennedy,
    author = {Kennedy, K. F.},
    title = {A Method for Analyzing Spread, Elongation and Bulge in Flat Rolling},
    journal = {J.~Eng. Ind.},
    volume = {109},
    number = {3},
    pages = {248--256},
    year = {1987},
    doi = {10.1115/1.3187126},
}

@phdthesis{Nikhaily,
  title={Metal Flow Models for Shape Rolling},
  author={El-Nikhaily, A. E. G. },
  school={Technical University of Aachen},
  year={1979}
}

@article{applied1961formula,
  title={Formula for `Spread' in Hot Flat Rolling},
  author={{Applied Mechanics Group} and Sparling, L. G. M. },
  journal={Proc. Inst. Mech. Eng.},
  volume={175},
  number={1},
  pages={604--640},
  year={1961},
  publisher={SAGE Publications Sage UK: London, England},
  doi={10.1243/PIME_PROC_1961_175_043_02}
}

@article{lahoti1974hill,
  title={On {H}ill's general method of analysis for metal-working processes},
  author={Lahoti, G. D. and Kobayashi, Shiro},
  journal=IJMS,
  volume={16},
  number={8},
  pages={521--540},
  year={1974},
  publisher={Elsevier},
  doi={10.1016/0020-7403(74)90018-6}
}

@article{oh1975approximate,
  title={An approximate method for a three-dimensional analysis of rolling},
  author={Oh, S. I. and Kobayashi, Shiro},
  journal=IJMS,
  volume={17},
  number={4},
  pages={293--305},
  year={1975},
  publisher={Elsevier},
  doi={10.1016/0020-7403(75)90010-7}
}

@article{carlsson1998contact,
  title={The contact pressure distribution in flat rolling of wire},
  author={Carlsson, B},
  journal=JMPT,
  volume={73},
  number={1-3},
  pages={1--6},
  year={1998},
  publisher={Elsevier},
  doi={10.1016/S0924-0136(97)00091-5}
}

@article{kazeminezhad2006pressure,
  title={Calculation of the rolling pressure distribution and force in wire flat rolling process},
  author={Kazeminezhad, M and {Karimi Taheri}, A},
  journal=JMPT,
  volume={171},
  number={2},
  pages={253--258},
  year={2006},
  publisher={Elsevier},
  doi={10.1016/j.jmatprotec.2005.06.070}
}

@article{kazeminezhad2008error,
  title={A study on the cross-sectional profile of flat rolled wire},
  author={Kazeminezhad, M and {Karimi Taheri}, A and {Kiet Tieu}, A},
  journal=JMPT,
  volume={200},
  number={1-3},
  pages={325--330},
  year={2008},
  publisher={Elsevier},
  doi={10.1016/j.jmatprotec.2007.09.029}
}

@article{kazeminezhad2005width,
  title={A theoretical and experimental investigation on wire flat rolling process using deformation pattern},
  author={Kazeminezhad, M and {Karimi Taheri}, A},
  journal={Materials \& Design},
  volume={26},
  number={2},
  pages={99--103},
  year={2005},
  publisher={Elsevier},
  doi={10.1016/j.matdes.2004.06.010}
}

@article{vallellano2008analysis,
  title={Analysis of deformations and stresses in flat rolling of wire},
  author={Vallellano, C and Cabanillas, P. A. and Garcia-Lomas, F. J.},
  journal=JMPT,
  volume={195},
  number={1-3},
  pages={63--71},
  year={2008},
  publisher={Elsevier},
  doi={10.1016/j.jmatprotec.2007.04.124}
}

@article{utsunomiya2001three,
  title={Three-dimensional elastic-plastic finite-element analysis of the flattening of wire between plain rolls},
  author={Utsunomiya, H and Hartley, Peter and Pillinger, Ian},
  journal={J. Manuf. Sci. Eng.},
  volume={123},
  number={3},
  pages={397--404},
  year={2001},
  doi={10.1115/1.1365158}
}

@article{kobayashi1978influence,
  title={Influence of rolling conditions on spreading in flat rolling of round wire},
  author={Kobayashi, M},
  year={1978},
  journal={J.~Jpn Soc. Technol. Plast.},
  volume={19},
  pages={630--637},
  note = {(in Japanese)}
}

@article{chitkara1966some,
  title={Some experimental results concerning spread in the rolling of lead},
  author={Chitkara, N. R. and Johnson, W},
  journal={J.~Basic Eng.},
  volume={88},
  number={2},
  pages={489--499},
  year={1966},
  publisher={ASME International},
  doi={10.1115/1.3645884}
}

@manual{MATLAB:R2024a_u1,
year = {2024},
author = {{MathWorks}},
title = {{MATLAB} version: 24.1.0.2568132 (R2024a) Update 1},
url = {https://www.mathworks.com}
}

@article{pesin2002mathematical,
  title={Mathematical modelling of the stress--strain state in asymmetric flattening of metal band},
  author={Pesin, A and Salganik, V and Trahtengertz, E and Cherniahovsky, M and Rudakov, V},
  journal=JMPT,
  volume={125},
  pages={689--694},
  year={2002},
  publisher={Elsevier},
  doi={10.1016/S0924-0136(02)00353-9}
}

@book{Semiatin1984FormabilityAW,
  title={Formability and workability of metals: plastic instability and flow localization},
  author={S. Lee Semiatin and John Joseph Jonas},
  year={1984},
  publisher={Am. Soc. Metals},
  isbn={978-0-87170-183-1}
}

@article{zhang2016relationship,
  title={Relationship between friction parameters in a {C}oulomb--{T}resca friction model for bulk metal forming},
  author={Zhang, Da-Wei and Ou, Hengan},
  journal={Tribology International},
  volume={95},
  pages={13--18},
  year={2016},
  publisher={Elsevier},
  doi={10.1016/j.triboint.2015.10.030}
}

@incollection{lin2008finite,
  title={Finite Element Simulation in Flat Rolling of Multi-Wire},
  author={Lin, Wei-Shin and Yang, Tung-Sheng and Hsieh, He-Jiun and Lin, Chun-Ming},
  booktitle={Advanced Design and Manufacture to Gain a Competitive Edge},
  pages={101--110},
  year={2008},
  publisher={Springer},
  doi={10.1007/978-1-84800-241-8_11},
}

@article{lee2018development,
  title={Development of width spread model for high carbon steel wire rods in flat rolling process},
  author={Lee, Kyunghun and Kim, Byungmin and Kim, Namjin},
  journal={Int. J. Mech. Eng. Robot. Res},
  volume={7},
  pages={343--347},
  year={2018},
  doi={10.18178/ijmerr.7.4.343-347},
}

@article{hwang2021influence,
  title={Influence of roll diameter on material deformation and properties during wire flat rolling},
  author={Hwang, Joong-Ki and Kim, Sung-Jin and Kim, Kee-Joo},
  journal={Applied Sciences},
  volume={11},
  number={18},
  pages={8381},
  year={2021},
  publisher={MDPI},
  doi={10.3390/app11188381},
}

\end{document}